\documentclass[twocolumn,aps,pra,amsmath,amssymb,superscriptaddress,showpacs]{revtex4}

\usepackage{graphicx}
\usepackage{graphics}
\usepackage{epsfig}
\usepackage[latin1]{inputenc}
\usepackage{latexsym}
\usepackage{amsmath}
\usepackage{amssymb}
\usepackage{enumerate}

\begin{document}


\title{Mermin inequalities for perfect correlations}


\author{Ad\'{a}n Cabello}
\email{adan@us.es}
\affiliation{Departamento de F\'{\i}sica Aplicada II,
Universidad de Sevilla, E-41012 Sevilla, Spain}
\author{Otfried G{\"u}hne}
\affiliation{Institut f\"ur Quantenoptik und Quanteninformation,
\"Osterreichische Akademie der Wissenschaften, A-6020 Innsbruck,
Austria}
\affiliation{Institut f\"ur Theoretische Physik,
Universit\"at Innsbruck, A-6020 Innsbruck, Austria}
\author{David Rodr\'{\i}guez}
\affiliation{Departamento de F\'{\i}sica Aplicada III,
Universidad de Sevilla, E-41092 Sevilla, Spain}

\date{\today}



\begin{abstract}
Any $n$-qubit state with $n$ independent perfect correlations is
equivalent to a graph state. We present the optimal Bell
inequalities for perfect correlations and maximal violation for all
classes of graph states with $n \le 6$ qubits. Twelve of them were
previously unknown and four give the same violation as the
Greenberger-Horne-Zeilinger state, although the corresponding states
are more resistant to decoherence.
\end{abstract}



\maketitle


\section{Introduction}


In 1989, Greenberger, Horne, and Zeilinger (GHZ) showed that no
local hidden variable (LHV) theory can assign predefined local
results which agree with the {\em perfect} correlations predicted by
quantum mechanics for separated measurements on $n \ge 3$ distant
sites on a system prepared in the $n$-qubit GHZ state \cite{GHZ89}.
Mermin converted the $n$-party GHZ proof into a violation of a
$n$-party Bell inequality \cite{Mermin90a}. The amount of the
violation of Mermin's inequalities, measured by the ratio ${\cal D}$
between the quantum value of the Bell operator and its bound in LHV
theories, grows exponentially with $n$. For a given $n$ (with $n$
odd), Mermin's inequality gives the maximal possible violation of
any $n$-party two-setting Bell inequality in quantum mechanics
\cite{WW01}.

Can we extend this result to other $n$-qubit states? The essential
ingredient for GHZ-type proofs and Mermin-type Bell inequalities is
that they require an $n$-qubit quantum state, which is a
simultaneous eigenstate of $n$ commuting local observables (i.e., a
stabilizer state). Any stabilizer state is, up to local rotations,
equivalent to a graph state \cite{VDM04} (i.e., a stabilizer state
whose generators can be written with the help of a graph
\cite{HEB04}). These states are essential in quantum error
correction \cite{Gottesman96} and one-way quantum computation
\cite{RB01}. For a small number of qubits, all classes of
nonequivalent graph states under single-qubit unitary
transformations are known \cite{HEB04}. For $3 \le n \le 6$, there
are thirteen different classes of graph states which are
nonequivalent to GHZ states. For a given $n$, some of them are more
robust against decoherence than the GHZ state \cite{DB04}.

Bell inequalities for graph states constitute a subject of intense
study \cite{DP97, GTHB05, Cabello05, SASA05, TGB06, Hsu06, B07,
W07}. However, the Mermin inequalities for most of them are unknown.
For a given state, the Mermin inequality is the Bell inequality such
that (I) the Bell operator is a sum of stabilizing operators of that
state (i.e., perfect correlations), and (II) the violation is
maximal. If the maximum is obtained for Bell operators with a
different number of terms, then we will choose the one with the
lowest number, since the other inequalities contain this inequality
and require more measurements. For some graph states, the Mermin
inequality is not unique due to the symmetries of the graph.

This definition is motivated by the relation between the original
GHZ proof \cite{GHZ89} and the Mermin inequality \cite{Mermin90a}.
The aim of this paper is to introduce the Mermin inequalities for
{\em all} graph states (or, equivalently, for all pure states with
$n$ independent perfect correlations) with $n<7$ qubits.

The graph state $|G\rangle$ is the unique $n$-qubit state that
satisfies $g_i |G\rangle = |G\rangle$, for $i=1,\ldots,n$, where
$g_i$ are the generators of the stabilizer group of the state,
defined as the set $\{s_j\}_{j=1}^{2^n}$ of all products of the
generators. The perfect correlations of the graph state are
\begin{equation}
\langle G | s_j | G \rangle = 1 \text{ for }j=1,\ldots,2^n.
\label{perfectcorrelations}
\end{equation}
The $g_i$'s are obtained with the help of a graph $G$. For instance,
the $n$-qubit GHZ state is associated to the star-shaped graph in
which qubit $1$ is connected to all the other qubits. This means
that $g_1 = X_1 \bigotimes_{i\neq 1}^{n} Z_i$ and $g_i = X_1 \otimes
Z_i$ for $i \neq 1$; $X_i$, $Y_i$, and $Z_i$ denote the Pauli
matrices acting on the $i$th qubit (see \cite{HEB04} for more
details).

There are many possible GHZ-like proofs for a given graph state
associated to a connected graph of $n \ge 3$ qubits. All of them
have the same structure. Any LHV theory assigning predefined values
$-1$ or $1$ to $X_i$, $Y_i$, and $Z_i$ in agreement with the quantum
predictions given by (\ref{perfectcorrelations}) must satisfy
\begin{equation}
s_j=1 \text{ for }j=1,\ldots,2^n. \label{qpredictions}
\end{equation}
However, if we choose a suitable {\em subset} of $q$ predictions
from the set (\ref{qpredictions}), and assume predefined values,
either $-1$ or $1$, then for some choices it happens that, at most,
only $p<q$ of these predictions are satisfied. For the remaining
$q-p$ quantum predictions, the prediction of the LHV theory is the
opposite: $s_j=-1$. This difference can be reformulated as a
violation of the Bell inequality
\begin{equation}
\beta \le 2p - q, \label{Bellforgraphs}
\end{equation}
where the Bell operator $\beta$ is the sum of the stabilizing
operators {\em of the chosen subset}. According to Eq.
(\ref{perfectcorrelations}), the graph state satisfies
\begin{equation}
\langle G | \beta | G \rangle = q.
\end{equation}
Therefore, $|G\rangle$ violates the inequality (\ref{Bellforgraphs})
by an amount ${\cal D} = q / (2p -q)$. For the GHZ proof with $n$
odd, the maximum contradiction, measured by $q/p$, and the maximum
violation of the Bell inequality, measured by ${\cal D}$, is
obtained when $q=2^{n-1}$ and $p=2^{n-2}+2^{(n-3)/2}$. This is
Mermin's inequality \cite{Mermin90a}. If we take a different subset
of stabilizing operators, then we can have a violation of a Bell
inequality, but usually not the maximum one.

Specifically, for a given graph state associated to a connected
graph of $n \ge 3$ qubits, if we consider the Bell operator
consisting of {\em the whole set of stabilizing operators}, then we
always have a violation of a Bell inequality \cite{GTHB05}, but not
the maximum one. A violation occurs because that Bell operator
contains a simpler Bell operator giving the maximum violation.


Why are we interested in those Bell inequalities with the maximum
${\cal D}$? ${\cal D}$ is the measure of nonlocality used in Refs.
\cite{Mermin90a, WW01, GTHB05}. For graph states and stabilizer Bell
inequalities, it is well defined, easily computable, and the two
practical measures of nonlocality, the resistance to noise and the
detection efficiency for a loophole-free Bell experiment, are
connected to ${\cal D}$.

(i) In actual experiments, instead of a pure state $|G\rangle$, we
usually have a noisy one, $\rho = {\cal V} \,|G\rangle \langle G|+
(1-{\cal V}) \openone /{2^n}$, where $\openone$ is the identity
matrix in the Hilbert space of the whole system. ${\cal D}$ is
related to the minimum value of ${\cal V}$ required to {\em actually
observe} a violation of the Bell inequality ${\cal V}_{\rm crit}$.
For graph states and stabilizer Bell inequalities, if ${\cal D}$
increases, then ${\cal V}_{\rm crit}$ decreases. Specifically, a
simple calculation gives that ${\cal V}_{\rm crit} = 1/{\cal D}$.

(ii) An open problem in fundamental physics is achieving a
loophole-free Bell experiment. A particularly important problem is
the {\em detection loophole} \cite{Pearle70}. ${\cal D}$ is related
to the minimum detection efficiency required for a loophole-free
Bell experiment $\eta_{\rm crit}$. For graph states and stabilizer
Bell inequalities, if ${\cal D}$ increases, then $\eta_{\rm crit}$
decreases. Specifically, for GHZ states and the Mermin inequality
with $n$ odd, $\eta_{\rm crit} = [2 + (\log 2 / \log {\cal D})]/4$
\cite{CRV07}.

(iii) In addition, ${\cal D}$ provides the relevant parameters of
the underlying GHZ-type proof: $q$ and $p$. Any GHZ-type proof can
be converted into an $n$-party quantum pseudotelepathy game in which
a team assisted with a graph state always wins, while a team with
only classical resources wins only with probability $p/q$
\cite{Cabello06a}. Therefore, the higher ${\cal D}$, the lower $p/q$
and the higher quantum advantage.


The knowledge of the Mermin inequalities for all graph states is
important for


(a) {\em Quantum information.} Graph states are essential for
quantum information tasks. Mermin inequalities are useful tools for
their experimental analysis. For instance, in recent experiments
preparing $6$-qubit graph states ${\cal V}$ is around $0.5$
\cite{LKSBBCHIJLORW05,LZGGZYGYP07,LGGZCP07}, thus Bell inequalities
with ${\cal D} >2$ are required to observe violation. We will show
that for all $6$-qubit graph states, Mermin inequalities have ${\cal
D} > 2$.

(b) {\em Nonlocality vs decoherence experiments.} For GHZ states,
${\cal D}$ increases exponentially with $n$ \cite{Mermin90a}.
However, GHZ states' entanglement lifetime under decoherence
decreases with $n$ \cite{DB04}. Therefore, a fundamental limitation
seemingly exists to observe macroscopic violations of Bell
inequalities with GHZ states. A natural question is: Does this
limitation also hold for other types of graph states? What happens
to those graph states whose lifetime under decoherence does {\em
not} decrease with $n$ \cite{DB04}? To answer these questions we
need to know how ${\cal D}$ scales with $n$ within a family of graph
states, and which graph states have higher ${\cal D}$.


\section{Mermin inequalities for graph states}


For each graph state, our task is to obtain, from all possible Bell
operators which are sums of stabilizing operators, those which
provide the highest violation. The exhaustive study for $n \ge 6$
becomes computationally difficult because the number of potential
Bell operators to test scales like $2^{2^n}$. However, if we
restrict our attention to Bell operators with the same symmetry as
the underlying graph, this investigation is still computationally
feasible for $n=6$.


\begin{figure}[htb]
\centerline{\includegraphics[width=\columnwidth]{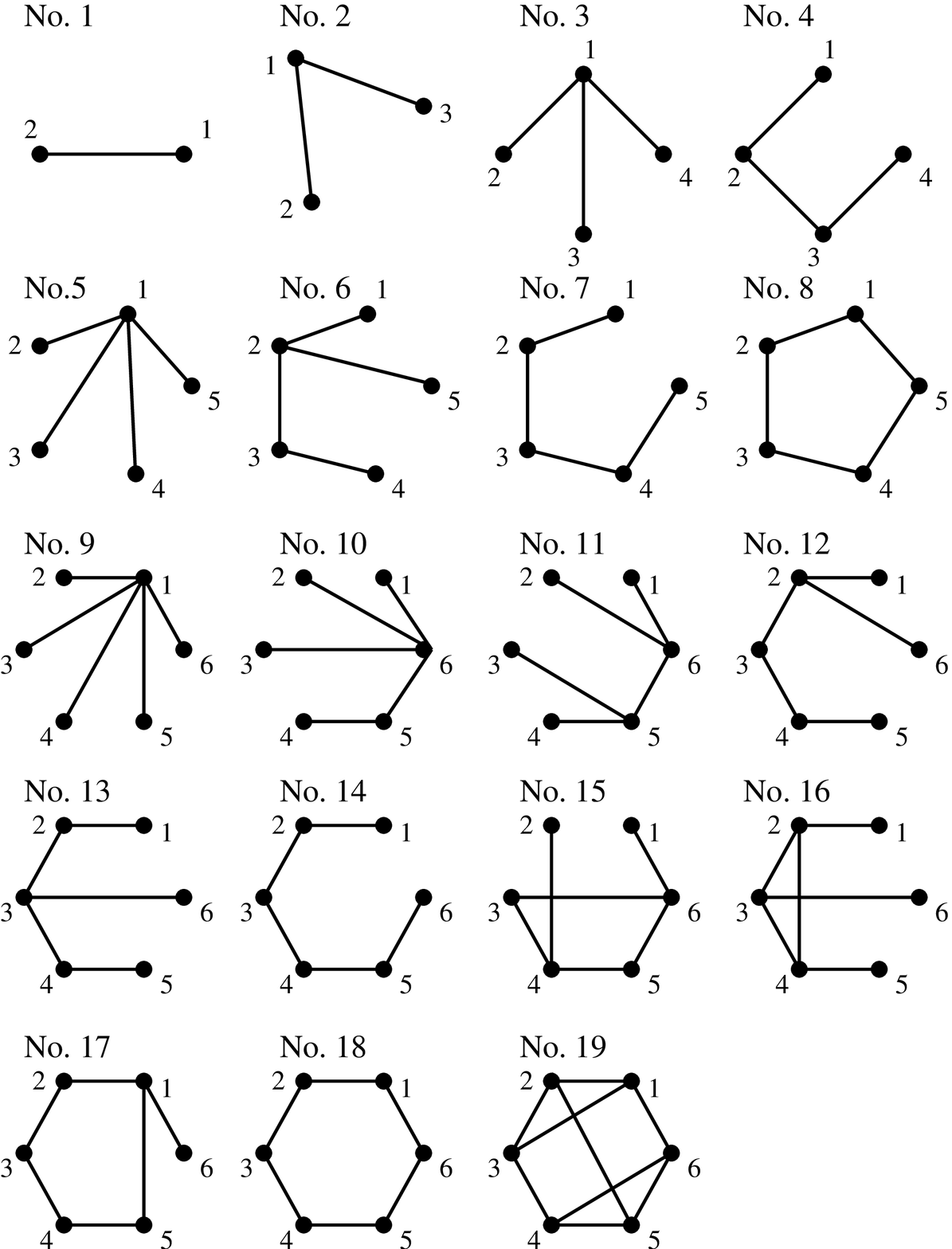}}
\caption{\label{Fig1} Graphs representing all classes of $n$-qubit
graph states, with $2\le n \le 6$, that are inequivalent under
single-qubit unitary transformations and graph isomorphism. The
figure is taken from Ref. \cite{HEB04}.}
\end{figure}


In Table \ref{TableI} we present all the Mermin inequalities for all
graph states with $2<n<6$ qubits. In Table \ref{TableII} we present
the Mermin inequalities possessing the same symmetry as the
underlying graph for all graph states with $n=6$ qubits. In both
tables we have followed the classification and the labeling of the
qubits of Fig. \ref{Fig1} (taken from Ref. \cite{HEB04}). LC$_n$
(RC$_n$) denotes the $n$-qubit linear (ring) cluster state
\cite{GTHB05}, $Y_5$ denotes the $5$-qubit graph state associated to
the graph ``$Y$'', $H_6$ the $6$-qubit graph state associated to the
graph ``$H$'', etc. The quantum prediction for each Bell operator
$\beta$ is $q$ (i.e., the number of terms of $\beta$); $p$ is the
maximum number of the $q$ perfect correlations that a LHV theory can
satisfy; ${\cal D}=\frac{q}{2p-q}$ is the violation of the Bell
inequality $\beta \le 2p-q$.


\begin{table*}[t]
\caption{\label{TableI}Mermin inequalities for all graph
states of $n<6$ qubits.}
\begin{ruledtabular}
{\begin{tabular}{lcccc}
${\rm Graph\;state}$ & $g_i$ & $\beta \le 2p-q$ & ${\rm Settings}$ & ${\cal D}$\\
\hline
2 (GHZ$_3$)
 & $g_1=X_1 Z_2 Z_3$
 & $g_1(\openone + g_2)(\openone + g_3) \le 2$
 & $2$-$2$-$2$
 & $2$\\
 & $g_i=Z_1 X_i$ for $i \neq 1$
 &
 &
 & \\
3 (GHZ$_4$)
 & $g_1=X_1 Z_2 Z_3 Z_4$
 & $g_1(\openone+g_2g_3+g_2g_4+g_3g_4) \le 2$ and $g_1 \rightarrow g_1 g_2$
 & $1$-$2$-$2$-$2$
 & $2$\\
 & $g_i=Z_1 X_i$ for $i \neq 1$
 & $g_1(\openone + g_i)(\openone + g_j) \le 2$ and $g_1 \rightarrow g_1 g_k$
 & $2$-$2(i)$-$2(j)$-$1(k)$
 & \\
4 (LC$_4$)
 & $g_1=X_1 Z_2$, $g_4=Z_3 X_4$
 & $(\openone + g_1)g_2(\openone + g_3) \le 2$ and $g_3 \rightarrow g_3 g_4$
 & $2$-$2$-$2$-$1$
 & $2$\\
 & $g_i=Z_{i-1} X_i Z_{i+1}$ for $i=2,3$
 & $(\openone + g_1)g_2(g_3 + g_4) \le 2$ and $g_3 \rightarrow g_3 g_4$
 & $2$-$2$-$1$-$2$
 & \\
 &
 & $g_i \rightarrow g_{i+1}$
 &
 & \\
5 (GHZ$_5$)
 & $g_1=X_1 Z_2 Z_3 Z_4 Z_5$,
 & $g_1(\openone + g_2)(\openone + g_3)(\openone + g_4)(\openone + g_5) \le 4$
 & $2$-$2$-$2$-$2$-$2$
 & $4$\\
 & $g_i=Z_1 X_i$ for $i \neq 1$
 &
 &
 & \\
6 ($Y_5$)
 & $g_1=X_1 Z_2$, $g_5=Z_2 X_5$
 & $g_2 \big[(\openone+g_1+g_5)(\openone+g_3+g_3g_4)+ (\openone+g_1 g_5) g_4\big]$
 &
 & \\
 & $g_2=Z_1 X_2 Z_5$
 & $+ (g_1+ g_5)g_3(\openone+g_4) \le 7$
 & $3$-$3$-$3$-$3$-$2$
 & $\frac{15}{7}$\\
 & $g_3=Z_2 X_3 Z_4$
 & $g_2 \rightarrow g_2 g_4$
 & $3$-$3$-$3$-$3$-$3$
 & \\
 & $g_4=Z_3 X_4$
 & $\beta \rightarrow g_4 \beta$ and $32$ nonsymmetric more
 &
 & \\
7 (LC$_5$)
 & $g_1=X_1 Z_2$, $g_5=Z_4 X_5$
 & $(\openone + g_1)\big[(\openone + g_2)g_3(\openone +g_4)+g_2g_4\big](\openone + g_5) \le 8$
 & $3$-$3$-$3$-$3$-$3$
 & $\frac{5}{2}$\\
 & $g_i=Z_{i-1} X_i Z_{i+1}$ for $i=2,3,4$
 &
 &
 & \\
8 (RC$_5$)
 & $g_i=Z_{i-1} X_i Z_{i+1}$
 & $\gamma + \sum_{i=1}^5 g_i g_{i+1} \le 9$
 & $3$-$3$-$3$-$3$-$3$
 & $\frac{7}{3}$\\
 &
 & $\gamma+ g_jg_{i+1} +g_ig_{i+2}+g_{i-1}g_{i+1}$
 &
 & \\
 &
 & $+g_{i-2}g_ig_{i+1}g_{i+2}+g_{i-2} g_{i-1}g_i g_{i+1} \le 9$
 & $3$-$3$-$3$-$3$-$3$
 & \\
 &
 & $\gamma=\frac{1}{2}\left[\prod_{i=1}^5(\openone + g_i)-\prod_{i=1}^5(\openone -g_i)\right]$
 &
 & \\
 &
 & and $105$ more
 &
 & \\
\end{tabular}}
\end{ruledtabular}
\end{table*}


\begin{table*}[t]
\caption{\label{TableII}Symmetric Mermin inequalities for all graph
states of $n=6$ qubits.}
\begin{ruledtabular}
{\begin{tabular}{lcccc}
${\rm Graph\;state}$ & $g_i$ & $\beta \le 2p-q$ & ${\rm Settings}$ & ${\cal D}$\\
\hline
9 (GHZ$_6$)
 & $g_1=X_1 Z_2 Z_3 Z_4 Z_5 Z_6$
 & $g_1 \big(\openone+\sum_{i \neq j \neq 1}g_i g_j+\sum_{i \neq j \neq k \neq l \neq 1} g_i g_j g_k g_l\big) \le 4$
 & $1$-$2$-$2$-$2$-$2$-$2$
 & $4$ \\
 & $g_i=Z_1 X_i$ for $i \neq 1$
 & and $g_1 \rightarrow g_1 g_2$
 & $1$-$2$-$2$-$2$-$2$-$2$
 & \\
10
 & $g_i=X_i Z_6$ for $i=1,2,3$
 & $(\openone+g_1) (\openone+g_2) (\openone+g_3) (\openone+g_5) g_6 \le 4$
 & $2$-$2$-$2$-$1$-$2$-$2$
 & $4$ \\
 & $g_4=X_4 Z_5$
 & $g_5 \rightarrow g_4 g_5$
 & $2$-$2$-$2$-$1$-$2$-$2$
 & \\
 & $g_5=Z_4 X_5 Z_6$
 & $(\openone+g_1) (\openone+g_2) (\openone+g_3) (g_4+g_5) g_6 \le 4$
 & $2$-$2$-$2$-$2$-$1$-$2$
 & \\
 & $g_6=Z_1 Z_2 Z_3 Z_5 X_6$
 & $g_5 \rightarrow g_4 g_5$
 & $2$-$2$-$2$-$2$-$1$-$2$
 & \\
11 (H$_6$)
 & $g_1=X_1 Z_6$, $g_2=X_2 Z_6$
 & $g_1 (\openone+g_2) (\openone+g_3) (\openone+g_4) (\openone+g_5) g_6 \le 4$
 & $1$-$2$-$3$-$3$-$3$-$2$
 & $4$ \\
 & $g_3=X_3 Z_5$, $g_4=X_4 Z_5$
 & $g_1 \leftrightarrow g_2$ (i.e., permute them)
 & $2$-$1$-$3$-$3$-$3$-$2$
 & \\
 & $g_5=Z_3 Z_4 X_5$, $g_6=Z_1 Z_2 X_6$
 & $(\openone+g_1) (\openone+g_2) g_3 (\openone+g_4) g_5 (\openone+g_6) \le 4$
 & $3$-$3$-$1$-$2$-$2$-$3$
 & \\
 &
 & $g_3 \leftrightarrow g_4$
 & $3$-$3$-$2$-$1$-$2$-$3$ & \\
12 (Y$_6$)
 & $g_1=X_1 Z_2$, $g_6=Z_2 X_6$
 & $(\openone+g_1) g_2 (\openone+g_3) g_4 (\openone+g_5) (\openone+g_6) \le 4$
 & $2$-$2$-$1$-$2$-$2$-$2$
 & $4$ \\
 & $g_2=Z_1 X_2 Z_3 Z_6$, $g_3=Z_2 X_3 Z_4$
 &
 &
 & \\
 & $g_4=Z_3 X_4 Z_5$, $g_5=Z_4 X_5$
 &
 &
 & \\
13 (E$_6$)
 & $g_1=X_1 Z_2$, $g_5=Z_4 X_5$
 & $(\openone+g_3+g_3 g_6)[(\openone+g_1)g_2+g_4(\openone+ g_5)$
 &
 & \\
 & $g_2=Z_1 X_2 Z_3$, $g_4=Z_3 X_4 Z_5$
 & $+(\openone+g_1)g_2 g_4(\openone+g_5)] \le 8$
 & $2$-$3$-$3$-$3$-$2$-$2$
 & $3$ \\
 & $g_3=Z_2 X_3 Z_4 Z_6$, $g_6=Z_3 X_6$
 & and $37$ more
 &
 & \\
14 (LC$_6$)
 & $g_1=X_1 Z_2$, $g_6=Z_5 X_6$
 & $(\openone+g_1) g_2 (\openone+g_3) (\openone+g_4) g_5 (\openone+g_6) \le 4$
 & $2$-$2$-$3$-$3$-$2$-$2$
 & $4$ \\
 & $g_i=Z_{i-1} X_i Z_{i+1}$ for $i=2,3,4,5$
 &
 &
 & \\
15
 & $g_1=X_1 Z_6$, $g_2=X_2 Z_4$
 & $(g_3+g_5)(\openone+g_1)(\openone+g_2)(\openone+g_4)(\openone+g_6)$
 &
 & \\
 & $g_3=X_3 Z_4 Z_6$, $g_5=Z_4 X_5 Z_6$
 & $+(\openone+g_3 g_5) (g_4+g_2 g_4+g_6+g_1 g_6) \le 16$
 & $3$-$3$-$3$-$3$-$3$-$3$
 & $\frac{5}{2}$ \\
 & $g_4=Z_2 Z_3 X_4 Z_5$, $g_6=Z_1 Z_3 Z_5 X_6$
 & and $6$ more
 &
 & \\
16
 & $g_1=X_1 Z_2$, $g_5=Z_4 X_5$
 & $g_3 (\openone+g_1+g_2+g_1g_2+g_4+g_5+g_4g_5)(\openone+g_6)$
 &
 & \\
 & $g_2=Z_1 X_2 Z_3 Z_4$, $g_4=Z_2 Z_3 X_4 Z_5$
 & $+(\openone+g_1)g_2(\openone+g_5+g_6)+g_4(\openone+g_5)(\openone+g_1+g_6)$
 &
 & \\
 & $g_3=Z_2 X_3 Z_4 Z_6$
 & $+(\openone+g_1)g_2g_4(\openone+g_5) \le 12$
 & $3$-$3$-$3$-$3$-$3$-$3$
 & $3$ \\
 & $g_6=Z_3 X_6$
 & and $3$ more
 &
 & \\
17
 & $g_1=X_1 Z_2 Z_6$
 & $[g_1 (\openone+g_2 g_5) (g_3+g_4)$
 &
 & \\
 & $g_2=Z_1 X_2 Z_3$, $g_5=Z_1 Z_4 X_5$
 & $+(\openone+g_1) (g_2+g_5) (\openone+g_3 g_4)] (\openone+g_6) \le 8$
 & $3$-$3$-$3$-$3$-$3$-$3$
 & $3$ \\
 & $g_3=Z_2 X_3 Z_4$, $g_4=Z_3 X_4 Z_5$
 & $\beta \rightarrow g_3 \beta$
 & $3$-$3$-$3$-$3$-$3$-$3$
 & \\
 & $g_6=Z_1 X_6$
 &
 &
 & \\
18 (RC$_6$)
 & $g_i=Z_{i-1} X_i Z_{i+1}$
 & $\sum_{i=1}^{64} s_i -\openone -\sum_{i=1}^6 g_i - g_1 g_3 g_5 - g_2 g_4 g_6 \le 19$
 & $3$-$3$-$3$-$3$-$3$-$3$
 & $\frac{55}{19}$ \\
19
 & $g_1=X_1 Z_2 Z_3 Z_6$, $g_4=Z_3 X_4 Z_5 Z_6$
 & $g_1 g_4+g_3 g_6+g_1 g_3 g_4 g_6+g_2 (g_4+g_6+g_4 g_6)$
 &
 & \\
 & $g_2=Z_1 X_2 Z_3 Z_5$, $g_5=Z_2 Z_4 X_5 Z_6$
 & $+g_5 (g_1+g_3 +g_1 g_3)+(g_2+g_5)(g_3 g_4+g_1 g_6+g_1 g_3 g_4 g_6)$
 &
 & \\
 & $g_3=Z_1 Z_2 X_3 Z_4$, $g_6=Z_1 Z_4 Z_5 X_6$
 & $+g_2 g_5 [g_1 g_4 (\openone+g_3+g_6)+g_3 g_6 (\openone+g_1+g_4)]\le 9$
 & $3$-$3$-$3$-$3$-$3$-$3$
 & $\frac{7}{3}$ \\
\end{tabular}}
\end{ruledtabular}
\end{table*}


Some of the inequalities in Tables \ref{TableI} and \ref{TableII}
were previously known. For the $n$-qubit GHZ states with $n$ odd, we
recover the original Mermin inequalities \cite{Mermin90a}. For the
$n$-qubit GHZ states with $n$ even, the original Mermin inequalities
are the sum of our two symmetric inequalities (the fist two
inequalities for the GHZ$_4$ in Table \ref{TableI} and the two
inequalities for the GHZ$_6$ in Table \ref{TableII}). Our
inequalities have the same violation as Mermin's, but only half of
the terms. For $n$ even, Ardehali proposed a method giving an
additional violation of $\sqrt{2}$ \cite{Ardehali92}. Ardehali's
inequalities do not use only perfect correlations. Ardehali's method
can be extended to other graph states \cite{GC08}. Mermin's
inequalities have been tested in the laboratory using $3$-
\cite{PBDWZ00} and $4$-qubit GHZ states \cite{SKKLMMRTIWM00,
ZYCZZP03}. Sources of $5$- \cite{ZCZYBP04} and $6$-qubit GHZ states
\cite{LKSBBCHIJLORW05, LZGGZYGYP07} already exist.


For the $4$-qubit cluster state (LC$_4$), the Mermin inequalities in
Table \ref{TableI} contain those introduced in \cite{Cabello05,
SASA05}. These inequalities have been recently tested in the
laboratory \cite{WARZ05, KSWGTUW05, VPMDB07}.

However, twelve of the Mermin inequalities in Tables \ref{TableI}
and \ref{TableII} are new. They include those for the RC$_5$,
important for quantum error correction codes \cite{Gottesman96}; for
the H$_6$, a universal resource for one-way quantum computation
recently prepared in the laboratory \cite{LZGGZYGYP07}; and for the
Y$_6$, which allows a demonstration of anyonic statistics in the
Kitaev model \cite{LGGZCP07, HRD07, PWSKPW07}.

A remarkable fact is, that four $6$-qubit graph states have the same
violation as the GHZ$_6$: the graph state no. 10, the $H_6$, the
$Y_6$, and the LC$_6$ (see Table \ref{TableII}). This is interesting
because these states are more resistant to decoherence than the
GHZ$_6$ \cite{DB04}. This proves that the nonlocality vs decoherence
ratio of GHZ states is not universal: there are states with similar
violations but that are more robust against decoherence.


\section*{Acknowledgments}


The authors thank H. J. Briegel, P. Moreno, and G. Vallone for
useful discussions. A.C. acknowledges support from the Spanish MEC
Project No. FIS2005-07689, and the Junta de Andaluc\'{\i}a
Excellence Project No. P06-FQM-02243. O.G. acknowledges support from
the FWF and the EU (OLAQUI, SCALA, and QICS).



\end{document}